\documentclass[aip]{revtex4-2}
\usepackage{amsmath,amssymb,amsfonts,graphicx}

\newcommand{\be}{\begin{equation}}
\newcommand{\ee}{\end{equation}}

\begin{document}
\title{Kuramoto model on the $D$-dimensional torus}

\author{Marcel Novaes}	
\affiliation{Instituto de F\'isica, Universidade Federal de Uberl\^andia, 38408-100, Uberl\^andia, MG, Brazil}

\begin{abstract}
	
We propose a generalization of the Kuramoto model of interacting oscillators in which the particles move on the surface of a $D$-dimensional torus. In contrast with the traditional one-dimensional version, this model has a first order phase transition. We establish its mean field dynamics by means of a multidimensional Ott-Antonsen ansatz, and show that synchronization arises from a saddle-node bifurcation, while the incoherent state is always stable. Our theoretical calculations are validated by numerical simulations.
	
\end{abstract}

\maketitle
	
\begin{quotation}
The Kuramoto model provides a simple description of interacting many-body systems that undergo a synchronization transition, and has been useful in their mathematical exploration. In the original version,  each particle is a phase oscillator and, as the coupling between then increases, the disorganized incoherent state loses stability by means of a transcritical bifurcation and the system spontaneously synchronizes in a second order phase transition. Here we generalize the model into a $D$-dimensional version, where each particle is characterized by $D$ phases. We analyze its mean field dynamics and show, for example, that this system experiences bistability and a first order phase transition.      
\end{quotation}

\section{Introduction}

The Kuramoto model is a mathematical representation of interacting oscillators that helps understand the onset of spontaneous synchronization in the time evolution of complex systems, a topic of relevance is multiple areas of science \cite{PR1,PR2,livro}. In its original and simplest version, each oscillator is represented by a phase variable $\theta_i(t)$, defined on the unit circle, that evolves with a natural frequency $\omega_i$ and also suffers the influence of a number of other oscillators, according to the equation of motion
\be \dot{\theta}_i =\omega_i+\frac{K}{N}\sum_{j\neq i}\sin(\theta_j-\theta_i).\ee
The natural frequencies are usually taken as independent identical random variables. When their distribution $g(\omega)$ is smooth, symmetric and unimodal, a second order phase transition occurs in the large-$N$ limit, with oscillators starting to synchronize as $K$ becomes larger than some critical value $K_c$. 

The intensity of the synchronization can be measured by the order parameter
\be z=re^{i\psi}=N^{-1}\sum_j e^{i\theta_j}.\ee
When motion is incoherent, the phases are effectively random and $r\approx 0$. Conversely, if all phases are very similar they interfere constructively, resulting in a sizeable value for $r$. When $g(\omega)$ is a Cauchy distribution with width $\Delta$, for example, one can show by means of the celebrated Ott-Antonsen ansatz \cite{OA}, that the stationary value of $r$ equals $\sqrt{1-2\Delta/K}$ for $K>K_c$.

This simple construction has been generalized in uncountable ways. Some versions display a first order phase transition, sometimes dubbed explosive synchronization, in which the stationary value of $r$ jumps abruptly from zero to a finite value. This phenomenon is usually accompanied by two characteristics. The first is bistability: there are two critical values $K_{c,1}$ and $K_{c,2}>K_{c,1}$; incoherence is certain for $K<K_{c,1}$, synchronization is certain for $K>K_{c,2}$, and both situations are stable in the  intermediate situation. The second is hysteresis: the order parameter grows abruptly at $K_{c,2}$ when $K$ is increased, and decreases abruptly at $K_{c,1}$ when $K$ is decreased. 

A first order phase transition has been observed under different situations. When higher-order interaction terms like\cite{higher1,higher2,higher3,arenas1,arenas2,higherPR} $\sin(2\theta_j-\theta_k-\theta_i)$ are incorporated, when $g(\omega)$ is non-standard\cite{pazo,bimodal}, when inertia terms proportional to $\ddot{\theta}$ are introduced,\cite{inertia0,inertia1} when interactions among oscillators follow the topology of some specific types of complex networks\cite{netw1,netw2,netw3}, or when the phase of each particle is replaced by a vector in a higher dimensional sphere\cite{sphereL,chandra1}. 

The Kuramoto model introduced in the present work has a first order phase transition without higher-order interactions, with a Cauchy $g(\omega)$, no inertia and a simple all-to-all connectivity. The phase of each unit is defined in a $D$-dimensional space, however we take this to be a torus and not a sphere. The choice of the torus has an advantage over the sphere: the multidimensional Ott-Antonsen ansatz leads to explicit equations of motion for the mean field approximation, whereas in the spherical case the effectiveness of the ansatz is somewhat limited\cite{chandra2,sphere0,barioni1,barioni2}.

Our system has an interesting dynamical peculiarity, a kind of giant hysteresis in which $K_{c,2}$ is infinite, so synchronization disappears abruptly when $K$ decreases through $K_{c}$, but incoherence is always stable and therefore synchronization never appears abruptly, no matter how much $K$ is increased (this behavior is also present in a model where the coupling constant depends on the order parameter \cite{infinite}).

Let us mention that a different kind of Kuramoto model on the $2$-torus has already been studied, defined as\cite{okeeffe,okeeffe2}
\be \dot{\theta}_i =\omega_i+\frac{K}{N}\sum_{j\neq i}\sin(\theta_j-\theta_i)\cos(\varphi_j-\varphi_i),\label{keefe1}\ee
\be \dot{\varphi}_i =v_i+\frac{J}{N}\sum_{j\neq i}\sin(\varphi_j-\varphi_i)\cos(\theta_j-\theta_i). \label{keefe2}\ee
It was actually interpreted as a population of swarmalators\cite{swarm0,liza1,swarm} on the circle, with $\varphi_i$ being considered the position of unit $i$ and $\theta_i$ its phase. This model has a very rich dynamics, but explicit equations of motion for the mean field approximation are not available because the Ott-Antonsen ansatz does not apply.

\section{The model}

We start by introducing a different model on the $2$-torus which is:
\be \dot{\theta}_i =\omega_i+\frac{K}{N^2}\left(\sum_{j\neq i}\sin(\theta_j-\theta_i)\right)\left(\sum_{j\neq i}\cos(\varphi_j-\varphi_i)\right),\ee
\be \dot{\varphi}_i =v_i+\frac{K}{N^2}\left(\sum_{j\neq i}\sin(\varphi_j-\varphi_i)\right)\left(\sum_{j\neq i}\cos(\theta_j-\theta_i)\right).\ee
We take the coupling constants to be equal, for simplicity. 

The main difference compared to (\ref{keefe1})-(\ref{keefe2}) is in the form of the coupling. Instead of summing, over all other units, the product of the $\theta$-interaction and the $\varphi$-interaction, we now multiply the averages of those two types of interactions, so the influences of other units on the dynamics of unit $i$ are no longer independent.

We can look at this model as Kuramoto-like, but with a dynamically evolving coupling in which the constant $K$ is effectively weighted, in the $\theta$ direction, by the average proximity of the $\varphi$ values, and vice-versa.

In our higher dimensional generalization, we do not interpret any of the degrees of freedom as a position. Instead, we consider each unit to be described by $D$ phases, $\{\theta_{i,1},\ldots,\theta_{i,D}\}$, and define a Kuramoto model on the $D$-dimensional torus by 
\be \dot{\theta}_{i,a} =\omega_{i,a}+K\langle\sin_a\rangle\prod_{b\neq a}\langle\cos_b\rangle,\ee
where $\langle\cdot\rangle$ denotes an average:
\be\label{medias} \langle f_a\rangle=\frac{1}{N}\sum_{j\neq i}f(\theta_{j,a}-\theta_{i,a}).\ee
Namely, the driving that unit $i$ feels along dimension $a$ equals an average sine in the same dimension, times a product of average cosines in all other dimensions.

It is now possible for the system to have partial synchronization along dimension $a$ if $\theta_{1,a}=\cdots\theta_{N,a}$, and total synchronization if all dimensions are synchronized. But notice that even at total synchronization, different dimensions may still differ in their final phase, i.e. we may have $\theta_{1,a}\neq \theta_{1,b}$. Synchronization along each dimension $a$ may be measured by a corresponding  mean field, a.k.a. order parameter
\be\label{zs} z_a=r_ae^{i\psi_a}=N^{-1}\sum_j e^{i\theta_{j,a}},\ee
or by 
\be x_a=r_a^2.\ee

Even for arbitrary $D$, some dynamical properties of the model may be derived immediately. For example, if all natural frequencies are zero (or are all equal and thus reduce to zero in an appropriate rotating frame) along a given dimension, then partial synchronization along that dimension is a linearly stable stationary state, just like for the usual Kuramoto model. 

Also, if a certain dimension $b$ is fully synchronized, it becomes invisible to the other dimensions, because in that case $\langle\cos_b\rangle=1$. The system then behaves effectively as if it were $(D-1)$-dimensional. In particular, if $D-1$ dimensions are synchronized, the dynamics along the remaining one follows the traditional Kuramoto model.

\begin{figure}
\includegraphics[scale=0.6]{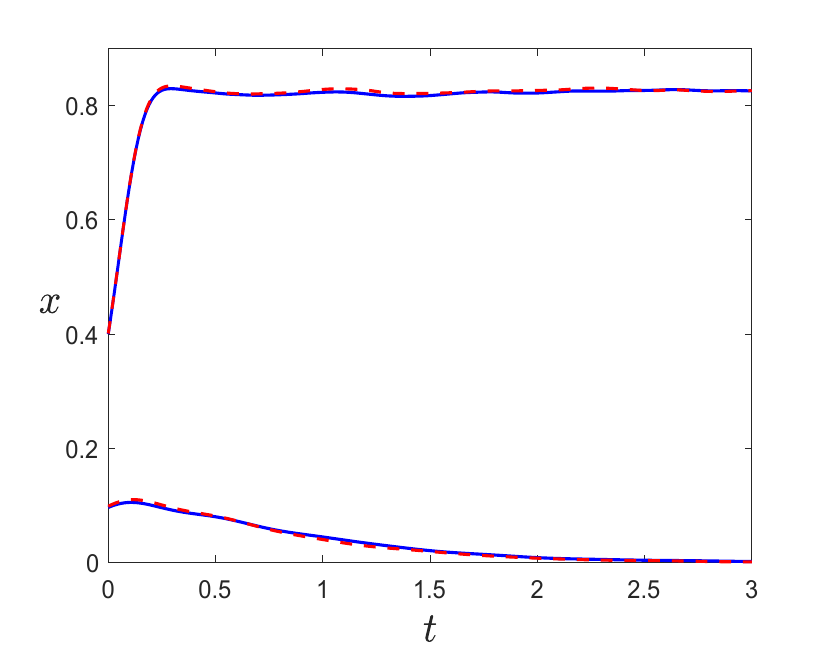}
\caption{Time evolution of the order parameters $x_1,x_2$ (depicted as solid and dashed lines, respectively) for a torus with two dimensions, $D=2$, with $g(\omega)$ a lorentzian centered at zero with width $1$ in both dimensions. The coupling is $K=15$, which gives a stationary value of approximately $0.84$ for $x(t)$. One initial condition leads to synchronization and the other leads to incoherence, as predicted by the theory.}
\end{figure}

\begin{figure}
\includegraphics[scale=0.6]{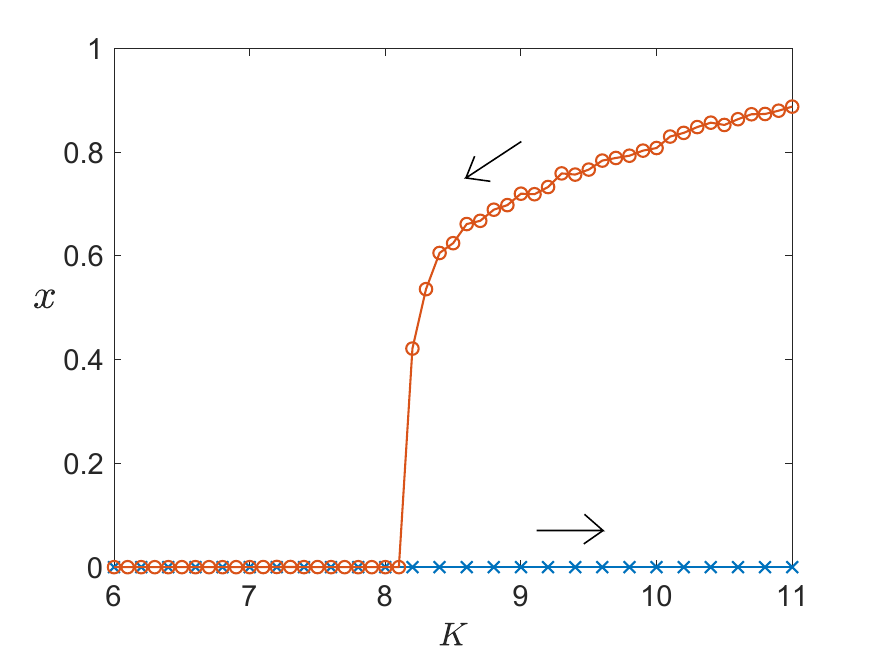}
\caption{Stationary values of the mean field $x$ as a function of $K$, with the same parameters as Fig.1. When $K$ is decreased, synchronization dies abruptly at $K_c=8$ (squares). When $K$ is increased, nothing happens (squares) because the incoherent state is always stable.}
\end{figure}

In Figure 1 and Figure 2, we show numerical simulations for $10^4$ oscillators on a $2$-dimensional torus. Natural frequencies are distributed in both dimensions according to a lorentzian $g(\omega)$ centered at zero with width $1$ (in practice we truncated the distribution at $|\omega|=15$). In Fig. 1 we see the time evolution of the order parameters $x_1$ and $x_2$, for two initial conditions, one above and one below the value of the unstable fixed point, which is $0.15$. The larger initial condition is attracted to synchronization and the lower one to incoherence, as predicted by the theory. The asymptotic value of the synchronized branch is close to the theoretical prediction of $0.84$. In Fig. 2 we see the asymptotic values of $x_1$ and $x_2$ as functions of $K$. A first order phase transition is observed when $K$ decreases, with $K_c=8$, but it is not present for increasing $K$. 

In order to gain further dynamical insight, we now turn to the limit $N\to\infty$ and resort to a multidimensional Ott-Antonsen ansatz.

\section{Probabilistic description}

In terms of the order parameters (\ref{zs}), we have $\langle\sin_a\rangle=r_a\sin(\psi_a-\theta_{i,a})$
and  $\langle\cos_b\rangle=r_b\cos(\psi_b-\theta_{i,b})$ (the sum in (\ref{medias}) actually does not include $j=i$, but in the $N\to \infty$ limit this is irrelevant). The equation of motion becomes
\be \dot{\theta}_{i,a} =\omega_{i,a}+ K r_a\sin(\psi_a-\theta_{i,a})\prod_{b\neq a}r_b\cos(\psi_b-\theta_{i,b}).\ee

In the continuum limit, we have the joint probability density $\rho(\vec{\theta},\vec{\omega},t)$ of having a $D$-dimensional phase vector with natural frequency $\vec{\omega}$ at position $\vec{\theta}$ at time $t$. This quantity must satisfy a continuity equation,
\be \frac{\partial \rho}{\partial t}+\sum_{a=1}^D\frac{\partial (\rho v^a)}{\partial \theta^a}=0,\ee
where
\be v_a=\omega_{a}+ K r_a\sin(\psi_a-\theta_{a})\prod_{b\neq a}r_b\cos(\psi_b-\theta_{b}).\ee

If we assume that different directions are uncorrelated,
\be \rho(\vec{\theta},\vec{\omega},t)=\prod_a \rho_a(\theta_a,\omega_a,t),\ee 
we can integrate the continuity equation over $D-1$ variables. For clarity, let us choose dimension $1$, for example, and integrate over $\theta_2,\ldots,\theta_D$. We obtain
\be \frac{\partial \rho_1}{\partial t}+\sum_{a=1}^D\int d\theta_2\cdots d\theta_D \frac{\partial (\rho v_a)}{\partial \theta_a}=0.\ee 

Every term in the above sum with $a\neq 1$ contains the term
\be \int d\theta_a \frac{\partial (\rho v_a)}{\partial \theta_a}.\ee By the fundamental theorem of calculus and periodicity around the circle, this quantity vanishes. On the other hand, the term with $a=1$ is
\be \frac{\partial}{\partial \theta_1}\int d\theta_2\cdots d\theta_D (\rho_1\cdots\rho_D v_1).\ee Velocity $v_1$ depends on other phases via order parameters. The integral to be done is simply
\be \int d\theta_b \rho_b \cos(\psi_b-\theta_b)=r_b.\ee

Therefore, the continuity equation along dimension $a$ is given by
\be \frac{\partial \rho_a}{\partial t}+\frac{\partial (\rho_a u_a)}{\partial \theta_a}=0,\ee
with the effective velocity
\be u_a=\omega_a+Kr_a\sin(\psi_a-\theta_a)\left(\prod_{b\neq a}r_b^2\right).\ee

\section{Multidimensional Ott-Antonsen ansatz}

Decomposing each density in a Fourier series,
\be \rho_a(\theta_a,\omega_a,t)=\frac{g(\omega_a)}{2\pi}\left(1+\sum_{n>0}c_{n,a}(\omega_a,t) e^{-in\theta_a}+c.c\right),\ee
the dynamical equations for the coefficients become
\be \frac{\dot{c}_{n,a}}{n}=i\omega_a c_{n,a}+\frac{K}{2}\left(z_ac_{n-1,a}-\bar{z}_ac_{n+1,a}\right)\left(\prod_{b\neq a}r_b^2\right).\label{cn}\ee

We now introduce a multidimensional Ott-Antonsen ansatz, namely that for each dimension $a$ there is a complex number $\alpha_a(\omega_a,t)$ such that the Fourier coefficents are all given by
\be c_{n,a}=\alpha_a^n.\ee
Substituting this expression into (\ref{cn}) leads to a dynamical equation for $\alpha_a$:
\be \dot{\alpha}_a=i\omega_a \alpha_a+\frac{K}{2}(z_a-\bar{z}_a\alpha_n^2)\left(\prod_{b\neq a}r_b^2\right),\ee
where 
\be z_a=\int g(\omega_a)\alpha_a d\omega_a.\ee

In the case when each $\omega_a$ has a Cauchy distribution centered around 0 with width $\Delta_a$, the above integral can be performed using residue theory (assuming $g$ to be analytic) to arrive at $z_a=\alpha_a(i\Delta)$. This leads to coupled dynamical equations for the order parameters,
\be \dot{z}_a=-\Delta_a z_a+\frac{K}{2}z_a(1-|z_a|^2)\left(\prod_{b\neq a}r_b^2\right).\ee
The dynamics of the phases is trivial, $\dot{\psi}_a=0$. Assuming for simplicity that $\psi_a=0$, we are left with the equations for the moduli,
\be \dot{r}_a=-\Delta_a r_a+\frac{K}{2}r_a(1-r_a^2)\left(\prod_{b\neq a}r_b^2\right).\ee
Changing variables to $x_a=r_a^2$, the final result for the mean field dynamics is 
\be \dot{x}_a=-2\Delta_a x_a+K(1-x_a)\left(\prod_{b}x_b\right).\ee
Notice how the last product now contains all $x$.

\section{Mean field dynamics}

\subsection{Stability of the incoherent state}

It is clear from these equations that if one of the $x$ is identically zero, this is enough to make all of them converge to zero. So incoherence in one dimension induces incoherence in all dimensions.

Moreover, due to the highly nonlinear interaction term, the linearized dynamics for small $x$ is simply $\dot{x}_a\approx -2\Delta_a x_a$, so the incoherent state is stable for all values of the coupling constant $K$, in marked contrast with the traditional Kuramoto model, in which the incoherent state becomes unstable for a critical value of $K$. As a consequence, the present system never evolves from strict incoherence to coherence (except in the exceptional case that all but one of the dimensions are already synchronized).

\subsection{Cooperation}

Since $\frac{\partial \dot{x}_a}{\partial x_b}>0$, the system is always cooperative, which implies there are no stable periodic orbits and almost all initial conditions will converge to attracting fixed points. 

Also, if $\Delta_a=\Delta_b$ it follows that 
\be \frac{1}{x_a-x_b}\frac{d}{dt}(x_a-x_b)<0,\ee
so if $x_a(0)>x_b(0)$ then $x_a(t)>x_b(t)$ for all $t>0$. Therefore, mean field trajectories of dimensions with the same frequency dispersion never cross, and in fact must converge towards each other. This can also be seen from the stationarity condition 
\be 2\Delta_a x_a=K(1-x_a)\left(\prod_{b}x_b\right)\ee
since it implies
\be \frac{\Delta_a x_a}{\Delta_b x_b}=\frac{1-x_a}{1-x_b},\ee
and it follows that, in the stationary regime, $x_a=x_b$ if $\Delta_a=\Delta_b$. 

\subsection{The saddle-node is generic}

Defining $X=\prod_a x_a$ and multiplying all equations of motion in the stationary regime, we get 
\be \frac{\prod_a (2\Delta_a+KX)}{X^{D-1}}=K^D.\label{geral}\ee
Let $F(X)$ denote the left hand side of this equation. On the half-line $X>0$, this function is positive, and it diverges on both sides, $F(0^+)=\infty$ and $F(\infty)=\infty$. Its second derivative is always positive, $F''(X)>0$, hence it has a single minimum and can equal the right hand side at most twice. 

This shows that, apart from the incoherent state, there can be at most two more stationary states. Therefore, for any value of $D$ the coupling $K$ has a single critical value, which is the locus of a saddle-node bifurcation, from which emerge a pair of stable/unstable solutions.
\subsection{Critical values}

The critical values of $K$ and $X$ correspond to the point where $F(X_c)=K_c^D$ and $F'(X_c)=0$. Taking the logarithmic derivative for convenience, we arrive at
\be \sum_a \frac{K_cX_c}{2\Delta_a+K_cX_c}=D-1.\ee
But this is equivalent to the very simple characterization
\be \sum_a x_{a,c}=D-1,\label{soma}\ee
which gives us an interesting relation tying together the critical values of all order parameters: they must always add up to $D-1$. In particular, the average value of a critical order parameter is $\bar{x}_c=1-1/D$, so synchronization becomes increasingly difficult as $D$ grows.

For any set of dispersions, the quantity $\gamma=K_cX_c$ can be obtained numerically from the equation $\sum_a \frac{\gamma}{2\Delta_a+\gamma}=D-1$. Then the critical value of each order parameter can be found as $x_{a,c}=\frac{\gamma}{2\Delta_a+\gamma}$, from which we can compute $X_c$ and then $K_c=\gamma/X_c$. 

\subsection{Large $K$}

For large $K$, we can linearize the stationary equations around the fully synchronized state. This shows that the stable fixed point is approximately given by 
\be x_{a,s}\approx 1-\frac{2\Delta_a}{K},\label{stable}\ee
regardless of dimension. On the other hand, linearizing around the incoherent state we find that the unstable fixed point is approximately given by
\be x_{a,u}\approx \frac{1}{\Delta_a}\left(\frac{2}{K}\prod_{b=1}^D\Delta_b\right)^{\frac{1}{D-1}}.\label{unstable}\ee

\section{Special cases}

In the following, we discuss a few special cases of the model, for which it is possible to find exactly the critical value $K_c$ and the values of the order parameters at the bifurcation.

\subsection{Two dimensions, $D=2$}

The stationary equations are 
\be 2\Delta_1 =Kx_2(1-x_1),\ee
\be 2\Delta_2 =Kx_1(1-x_2).\ee

In the simplest case of equal dispersions, $\Delta_1=\Delta_2=\Delta$, we have $x_1=x_2=x$ and
\be x^2-x+\frac{2\Delta}{K}=0,\ee
with solutions
\be x=\frac{1\pm\sqrt{1-8\Delta/K}}{2},\label{xK}\ee
showing that $K_c=8\Delta$ in this case. The larger/smaller value of $x$ is the stable/unstable state. The stable branch grows to 1 as $K$ increases, while the unstable one vanishes as $K$ increases, in agreement with (\ref{stable}) and (\ref{unstable}). Eq.(\ref{xK}) is verified in Figures 1 and 2.

We mention again that the incoherent state is always stable, so there is bistability for $K>K_c$. Starting from the synchronized state for large $K$ and decreasing $K$, the system will experience a sudden transition to incoherence at $K_c$ (but not the other way around).

With different dispersions, we get
\be Kx_1^2-(K-2\Delta_1+2\Delta_2)x_1-2\Delta_2=0,\ee
and a symmetrical version for $x_2$. Setting the discriminant to zero gives us the critical coupling as
\be \sqrt{K_c}=\sqrt{2\Delta_1}+\sqrt{2\Delta_2}.\ee
At this point, the corresponding values of the order parameters are
\be x_{1,c}=\frac{\sqrt{\Delta_2}}{\sqrt{\Delta_1}+\sqrt{\Delta_2}},\quad x_{2,c}=1-x_{1,c}.\ee

\subsection{All $\Delta$ equal}

If all $\Delta$ are equal, then all stationary $x$ must be equal, and we have a single equation,
\be Kx^{D}-Kx^{D-1}+2\Delta=0.\label{allD}\ee
Relation (\ref{soma}) implies that $x_c=1-1/D$, which corresponds to a critical coupling of
\be K_c=2\Delta \frac{D^D}{(D-1)^{D-1}}.\ee

\subsection{All but one $\Delta$ equal}

Now let us suppose $\Delta_1$ stands alone and $\Delta_2=\cdots=\Delta_{D}$. Then $x_2=\cdots=x_{D}=x$. Let $x_1=y$. Stationary equations are
\be K x^{D-1}y (1 - y) = 2\Delta_1 y,\ee
\be K x^{D-1}y (1 - x) = 2\Delta_2 x,\ee
from which one can show that
\be (\Delta_1 - \Delta_2)y_c^2 + D\Delta_2 y_c - (D-1)\Delta_2 = 0.\ee
Equation (\ref{soma}) implies that $y_c=(D-1)(1-x_c)$, and the critical coupling is
\be K_c=\frac{2\Delta_1}{x_c^{D-1}(1-y_c)}.\ee

\subsection{Two different $\Delta$ for even $D$}

Let $D=2M$ and suppose there are only two different dispersion values, $\Delta_1=\cdots=\Delta_{M}=\Delta$ and $\Delta_{M+1}=\cdots=\Delta_{2M}=\eta\Delta$. Let $x_1=\cdots=x_{M}=x$ and $x_{M+1}=\cdots=x_{2M}=y$. Stationarity equations are
\be K x^{M}y^M(1 - x) = 2\Delta x,\label{two}\ee
\be K x^{M}y^M(1 - y) = 2\eta\Delta y.\ee
In this case $y_c=2-\frac{1}{M}-x_c$ and 
\be M(\eta-1)x_c^2-[(3M-1)\eta-M+1)]x_c+(2M-1)\eta=0.\ee

\section{Conclusion}

We have introduced a new Kuramoto model in which the phases of the units are replaced by $D$-dimensional variables, however not on the surface of a sphere as traditionally done but on a torus. If the natural frequencies in different dimensions are chosen independently from Cauchy distributions, this leads to a mean field theory based on a multidimensional version of the Ott-Antonsen ansatz that produces explicit equations of motions for the order parameters. 

Using these equations, we proved that the incoherent state is always stable, and that incoherence in one dimension induces incoherence in the other ones. Also, there is a single saddle-node bifurcation at a critical value of the coupling constant, producing a first order phase transition in which synchronization vanishes abruptly when the coupling is lowered below the critical value (spontaneous synchronization out of incoherence does not happen). We also found approximations for the stable and unstable fixed points for large $K$.

The critical values of the order parameters satisfy a surprisingly simple sum rule, $\sum_a x_a=D-1$, characterizing the locus of the bifurcation as a hyperplane in parameter space. Finding the corresponding critical value of the coupling $K_c$ explicitly is not possible in general, although we indicate a numerical route for accomplishing this. We explored some particular cases of the model where determining critical values reduces to solving a quadratic equation. 

Further generalizations of this model will be investigated in future works. Kuramoto models on the sphere can be adapted to many different situations, with rich dynamical behavior\cite{mobius,sphere2,sphere3}, and it would be interesting to see if that is the case for our version on the torus.

\end{document}